\documentstyle[aps,epsfig,floats,tighten,12pt]{revtex}

\begin{document}
\title
{\bf Strange Magnetism and the Anapole Structure of the Proton}

\author{ 
R.~Hasty$^{2}$, A.~M.~Hawthorne-Allen$^{5}$, T.~Averett$^{9}$
D.~Barkhuff$^{4}$, D.~H.~Beck$^{2}$, E.~J.~Beise$^{3}$,  
A.~Blake$^{1}$, H.~Breuer$^{3}$,
R.~Carr$^{1}$,  S.~Covrig$^{1}$, A.~Danagoulian$^{2}$,
G.~Dodson$^{4}$, K.~Dow$^{4}$, 
M.~Farkhondeh$^{4}$, 
B.~W.~Filippone$^{1}$, J.~Gao$^{1}$, M.~C.~Herda$^{3}$,   
T.~M.~Ito$^{1}$, C.~E.~Jones$^{1}$, 
W.~Korsch$^{6}$, K.~Kramer$^{9}$,
S.~Kowalski$^{4}$, P.~Lee$^{1}$, 
 R.~D.~McKeown$^{1}$, B.~Mueller$^{7}$, 
M.~Pitt$^{5}$, J.~Ritter$^{5}$, J.~Roche$^{9}$,
V.~Savu$^{1}$, D.~T.~Spayde$^{3}$, R.~Tieulent$^{3}$, E.~Tsentalovich$^{4}$,
S.~P.~Wells$^{8}$, B.~Yang$^{4}$, and T.~Zwart$^{4}$
}

\address{
$^{1}$ Kellogg Radiation Laboratory, California Institute of Technology
Pasadena, CA 91125, USA \\
$^{2}$ Department of Physics, University of Illinois at Urbana-Champaign, 
Urbana, Illinois 61801 \\
$^{3}$ Department of Physics, University of Maryland, College Park, Maryland 
20742 \\
$^{4}$ Bates Linear Accelerator Center, Laboratory for Nuclear Science
and Department of Physics,\\
Massachusetts Institute of Technology, Cambridge, Massachusetts 02139 \\
$^{5}$ Department of 
Physics, Virginia Polytechnic Institute and State University, Blacksburg, 
VA 24061-0435  \\
$^{6}$ Department of Physics and
Astronomy, University of Kentucky, Lexington, KY  40506  \\
$^{7}$ Physics Division, Argonne National Laboratory, Argonne, IL 60439, USA\\
$^{8}$  Department of Physics, Louisiana Tech University, Ruston, 
LA 71272, 
USA  \\
$^{9}$ Department of Physics, College of William and Mary,
Williamsburg, VA 23187, USA \\
}

\maketitle

\eject

The violation of mirror symmetry in the weak force provides 
a powerful tool to study the internal structure of the proton. 
Experimental results have been obtained that
address the role of strange quarks in generating 
nuclear magnetism. The measurement reported here provides an
unambiguous constraint on strange quark contributions to
the proton's magnetic moment through the electron-proton weak
interaction.  We also report evidence for the 
existence of a parity-violating electromagnetic
effect known as the anapole moment of the proton.
The proton's anapole moment is not yet well understood 
theoretically, but it could have important implications for 
precision weak interaction studies in atomic systems such as cesium.

\vskip 12pt
\eject

In 1933 the German physicist Otto Stern discovered that the 
magnetism of the proton was anomalously large, 
a factor of three larger than expected from the basic theory of 
quantum mechanics. This experiment turned out to be the first glimpse 
of the internal structure of the constituents
of the atomic nucleus, and a tantalizing hint at the existence of
quarks. Widespread applications of the proton's magnetic properties,
such as the magnetic resonance imaging (MRI) techniques 
used in biology and medicine,
have been developed despite a lack of fundamental understanding
of the basic dynamics that generates this magnetism.
After the key discovery
of internal structure in the proton in a high energy electron 
scattering experiment at the 
Stanford Linear Accelerator Center in the late 
1960's~{\it (1)},
the theory of Quantum Chromodynamics (QCD), which describes the interaction
between quarks and the gluons that bind the quarks into the
atomic nuclei observed in the periodic table, was developed.
Despite almost 30 years of intense theoretical effort, QCD has 
been unable to produce numerical predictions for the basic properties
of nucleons such as their degree of magnetism.

In 1988, Kaplan and Manohar~{\it (2)} proposed
that the study of the weak magnetic force (analogous to the usual magnetic
force associated with electromagnetism) would allow a separation of the
proton's magnetism into the three contributing flavors of quarks: up, down
and strange.
Measuring the contribution from
strange quark-antiquark pairs is of special interest because it 
relates directly  to  the ``sea'' of virtual quark-antiquark pairs
in the proton, a phenomenon predicted by QCD.
In 1989, it was noted that the weak magnetic force
could be isolated using its unique property of 
lack of mirror symmetry, or parity violation {\it (3)}. 
The  basic idea was to study the preference for the
proton's interaction with electrons that have spin counterclockwise,
over those with clockwise spin, relative to their direction 
of travel. We report data obtained using this method that, 
when combined with our previously published results~{\it (4,5)},
allow the first unambiguous determination of the proton's weak
magnetism. We also report a measurement of a parity-violating,
time-reversal-even 
electromagnetic contribution to proton structure, referred to
in the literature as the proton's anapole moment.

The weak magnetic moment of the proton, designated as $\mu_Z$, 
is associated with the interaction due to exchange of the weak neutral
$Z$ particle.  The weak and electromagnetic forces are related
through the mixing angle $\theta_W$, a fundamental parameter 
of the standard model of electroweak interactions that relates
the $Z$~boson to its charged counterparts, the 
$W^{\pm}$ (we use $\sin^2\theta_W$=$0.23117 \pm 0.00016$~{\it (6)} in our 
analysis).
As a result, one can write $\mu_Z$ as 
\begin{eqnarray}  
\mu_Z =  (\mu_p-\mu_n) - 4 \sin^2\theta_W \> \mu_p - \mu_s
\end{eqnarray}  
where $\mu_{p(n)}$ are the usual magnetic moments of the proton (neutron)
(2.97$\mu_N$ and $-$1.91$\mu_N$, respectively, where 
$\mu_N=e\hbar /2m_p=3.152451238(24) \times 10^{-14}$ MeV-T {\it (6)}).
The individual contributions of up, down  and strange 
quark-antiquark pairs to the proton's magnetic moment,
$\mu_u$, $\mu_d$, and $\mu_s$, are defined by the relation
$\mu_p = {2\over 3}\mu_u - {1\over 3}\mu_d - {1\over 3}\mu_s \, $.
The neutron's magnetic moment is similarly constructed, interchanging
only $\mu_u$ and $\mu_d$.  Thus, a measurement of $\mu_Z$ 
combined with the known $\mu_p$ and $\mu_n$, provides the third
observable required to uniquely determine the $u$, $d$ and $s$
quark pieces of the proton's magnetic moment.
The study of parity-violating electron-nucleon ($e$-$N$)
scattering enables a determination of the weak magnetic form factor 
$G_M^Z(Q^2)$, and as a result the equivalent strange piece $G_M^s(Q^2)$, the
momentum-dependent counterpart of $\mu_Z$, where $Q^2$ is the relativistic
four-momentum transfered to the proton by the electron. The $Q^2$ dependence
of these quantities is sensitive to their spatial variation inside the proton.

Whereas the weak magnetism discussed above is a 
vector $e$-$N$ interaction, an axial vector $e$-$N$ coupling 
also exists, which is related to the proton's intrinsic spin.
The parity violating $e$-$p$ interaction depends on both of these quantities,
and it is essential to determine the axial vector $e$-$N$ form factor 
$G_A^e$ in order to reliably extract $G_M^s$.
In general, $G_A^e$ may be written as
$G_A^e = G_A^Z + \eta F_A + R^e$,   
where $G_A^Z$ is the contribution from a single $Z$-exchange
as would be measured in neutrino-proton elastic scattering (Fig.~1A),
$F_A$ is the nucleon anapole moment~{\it (7)},
and $R^e$ is a radiative correction (Fig.~1C). 
The constant $\eta$ is 
$\frac{8\pi\sqrt{2}\alpha}{1-4\sin^2\theta_W}=3.45$
where $\alpha$ is the fine structure constant. 
The anapole moment $F_A$ is identified
as the parity-violating coupling of a photon to the nucleon (Fig.~1B)
and is expected to be the largest of a class of
higher order interactions, or radiative corrections{\it (8)}.
It can arise as a result of, for example, a weak interaction between
two quarks inside the nucleon.
It is analogous to the nuclear anapole moment recently measured in
atomic cesium~{\it (9)},
which is enhanced by parity violating interactions between nucleons
in the cesium  nucleus~{\it (10)}.
Technically, the  theoretical separation of the anapole moment
from other radiative corrections $R^e$ is gauge dependent,
so the terminology associated with
these amplitudes varies in the literature, and $F_A$ itself 
is not cleanly defined theoretically.
However, because the anapole
moment is an electromagnetic interaction, it does not
contribute to neutrino scattering, and is unique to parity-violating
interactions with charged particles like electrons. Here we
identify the observed difference as due to the anapole 
contribution~{\it(11)}. 

We performed a measurement of the parity violating asymmetry in
the scattering of longitudinally polarized electrons from
neutrons and protons (nucleons) 
in deuterium using the SAMPLE apparatus at the 
MIT/Bates Linear Accelerator Center. The apparatus was that used in
our previously reported measurement on hydrogen~{\it (4,5)}.
In the deuterium experiment, the hydrogen target was replaced 
with deuterium, and borated polyethylene shielding was installed
around the target to reduce background from low energy neutrons
knocked out of the deuterium.  Combining this measurement with the
previously reported results allows an unambiguous determination of
both the axial $e$-$N$ form factor $G_A^e$ and the contribution
of strange quarks to the proton's magnetic form factor $G_M^s$. 
When presented in the context of the electron-quark couplings predicted
by the standard model for electroweak interactions, our result also places
new limits on the electron-quark axial couplings $C_{2u}$ and $C_{2d}$.

Parity violating electron scattering generally involves 
scattering of a longitudinally polarized electron beam from an
unpolarized target. The change in the count rate resulting from
reversal of the beam polarization indicates a parity nonconserving
effect. For elastic electron scattering from a proton, 
the asymmetry in the count rate can be written as 
\begin{equation}
A_{PV} = \left(-{G_F\over 4\pi\alpha\sqrt{2}}\right)
{ \varepsilon G_E^p G_E^Z + \tau G_M^p G_M^Z 
- \varepsilon^{\prime}\left(1-4\sin^2\theta_W\right) G_M^p G_A^e\over
\varepsilon\left(G_E^p\right)^2 + \tau\left(G_M^p\right)^2}\, .
\end{equation}
The three terms in $A_{PV}$ reflect the fact that it arises 
as a result of an interference between the electromagnetic and weak
interactions, where $G_{E,M}^p$ are the ordinary form factors
associated with the proton's charge and magnetic moment. 
An equivalent expression can be written for 
electron-neutron scattering. The kinematic factors
$\tau={Q^2\over 4 M_p^2}$, 
$\varepsilon = \left[1+2(1+\tau)\tan^2 (\theta_e/2)\right]^{-1}$ and 
$\varepsilon^{\prime}=\sqrt{(1-\varepsilon^2)\tau(1+\tau)}$
can be adjusted to enhance the 
relative sensitivity of the experiment to the three
contributions to $A_{PV}$.

In the SAMPLE experiment~{\it (12)}, the beam is generated from 
circularly polarized laser light incident on a bulk GaAs crystal,
resulting in an electron beam with an average polarization of 36\%.
The electrons are then accelerated in durations of 25~$\mu$sec
to an energy of 200~MeV before
encountering a liquid hydrogen (or deuterium) target.  
The polarization of the laser light is 
reversed at the accelerator repetition rate of 600~Hz. 
After encountering the hydrogen target, the scattered
electrons are detected in a large solid angle air {\v C}erenkov detector 
covering angles between 130$^\circ$ and $170^\circ$. Figure~2 
depicts 1 of 10 detector modules placed symmetrically about
the beam axis. All backward scattered electrons with energies above 20~MeV
travel faster than the speed of light in air and generate 
{\v C}erenkov light that is focussed onto a photomultiplier tube
by an ellipsoidal mirror. The integrated photomultiplier current
is proportional to the scattered electron rate, which is about
10$^8$ s$^{-1}$ during the beam pulse. A shutter in front of the 
photomultiplier tube allows measurement of background 
coming from neutrons and charged particles. 
As described in ref.~{\it(4)}, the pulse-height spectrum of the 
light-producing 
portion of the signal is characterized with dedicated runs using 
very low current beam. 
From these two pieces of information we determine the fraction of the 
signal due to the {\v C}erenkov yield, which varies from mirror to mirror,
but is typically 60\%. Several feedback systems are used to minimize
the dependence of the properties of the electron beam, such as position,
angle and energy, on its polarization state. 

The parity-violating asymmetry measured in SAMPLE is dominated by
scattering from the neutrons and protons rather than
the deuterium nucleus as a whole, and for our
incident electron energy of 200~MeV it can be written as
\begin{equation}
A_d = \left[ -7.27 + 1.78 G^e_A(T=1) + 0.75 G^s_M\right] \, {\mathrm ppm} 
\end{equation}
(where ``ppm'' stands for ``part per million'').
The analogous expression for elastic electron scattering on the proton is
\begin{equation}
A_p = \left[ -5.72 + 1.55G^e_A(T=1) + 3.49 G^s_M \right] \, {\mathrm ppm}.
\end{equation}
In these expressions we retain explicitly the isovector
(T=1, or proton $-$ neutron) component of $G_A^e$. The small isoscalar
component (proton + neutron) has been absorbed into the first
constant term. The combined measurements allow an independent 
experimental determination of both quantities.

The raw measured deuterium asymmetry is
$-$1.41$\pm$0.14~ppm, with a contribution due to 
helicity correlated beam parameters of 0.06$\pm$0.06~ppm.
The measured background contribution to the raw
asymmetry is 0.17$\pm$0.18~ppm, 
consistent with zero as expected. The dominant source of background
is from photoproduced neutrons, which is known to have a parity-violating 
asymmetry of less than 1.5\% of our measurement {\it (13)}. 
In our analysis we assume the background asymmetry is negligible.
Removal, on a detector-by-detector
basis, of all dilution factors (mostly due to the 
beam polarization but also from background yields) 
and application of small corrections for contributing physics 
processes yields a value for the physics asymmetry of
\begin{equation}
A_d = -6.79 \pm 0.64 \, {\mathrm (stat)} \pm 0.55 \, {\mathrm (sys)} \> {\rm ppm} \> .
\end{equation}
The dominant sources of systematic error are the beam parameter corrections 
(5\%), determination of the beam polarization (4\%), and determination of
the signal-to-background ratio (4\%). Uncertainties due to our lack of
knowledge of the deuterium nuclear structure are about 3\%~{\it (14)}.

Combining this measurement with the previously reported hydrogen 
asymmetry~{\it (5)}
and with the expressions in equations (3) and (4) leads to the two 
sets of diagonal bands in Fig.~3. The inner portion of each
band corresponds to the statistical error, and the outer portion
corresponds to statistical and 
systematic errors combined in quadrature.
Our best experimental value of $G_M^s$ is
\begin{equation}
G_M^s (Q^2=0.1) = 0.14 \pm 0.29 \, {\mathrm (stat)} 
\pm 0.31 \, {\mathrm (sys)} \> .
\end{equation}
In order to quote a result for the static contribution $\mu_s$
it is necessary to know the momentum dependence of $G_M^s$. 
We use the calculation of~{\it (15)}, resulting in 
\begin{equation}
\mu_s = [0.01 \pm 0.29 \, {\mathrm (stat)} \pm 0.31 \, {\mathrm (sys)}
\pm 0.07 \, {\mathrm (theor.)}] \> \mu_N
\end{equation}
where we have added an additional third uncertainty (theor.)
to account for error in the extrapolation coming from the range
of the theoretical prediction. Combining this result with the known
magnetic moments of the neutron and proton implies that the 
contribution of strange quarks to the proton's magnetic 
moment is $-0.1\pm 5.1$~\%.

The allowed region in Fig.~3 also provides a 
determination of
the isovector axial $e$-$N$ form factor. From a theoretical standpoint, 
the most uncertain contribution to $G^e_A$
is the anapole term. The dominant contribution is expected to come from
Fig.~1A, and at the kinematic conditions of the SAMPLE experiment
would result in a value of $G_A^e(T=1)$=$-$1.071$\pm$0.005~{\it (6,16)}.
This value is analogous to that which would be measured in neutrino
scattering. Anapole effects have been estimated to substantially 
reduce the magnitude of $G_A^e (T=1)$~{\it (17)},
and a recent update of that theoretical
treatment yields an expected value $G_A^e (T=1) = -0.83 \pm 0.26$~{\it (18)}.
This latter value is shown as the vertical band in Fig.~3.

Our experimental result is
\begin{equation}
G_A^e (T=1) = +0.22 \pm 0.45 \, {\mathrm (stat)}  \pm 0.39 \, {\mathrm (sys)} 
\end{equation}
indicating that the substantial modifications of $G_A^e$ predicted 
in~{\it (18)}
are not only present, but  with an even larger
magnitude than quoted.  The deviation of our measured
value of $G_A^e$ from $-1.071$ can be interpreted as a 
large anapole moment of the nucleon. 
In the case of cesium, the anapole moment is enhanced by collective 
nuclear effects that are not present for a single nucleon.  Understanding 
these higher order electroweak processes is essential to reliably interpret 
precision studies of atomic parity violation and other tests of the 
electroweak theory.  

If further theoretical work fails to produce agreement with this
experimental result, it may be necessary to consider the possibility
that the discrepancy is due to new physics not contained in the standard 
model of electroweak interactions. In that regard, we can recast our result
in terms of the electron-quark axial coupling parameters $C_{2u}$ and
$C_{2d}$~{\it (6)}.
Present limits on these quantities are from
the first parity violating electron scattering experiment~{\it (19)}
and from a measurement performed at the Mainz electron accelerator~{\it (20)}.
The form factor $G_A^e(T = 1)$ is related to the 
combination $C_{2u}-C_{2d}$, resulting
in our new experimental value of 
\begin{equation}
C_{2u} - C_{2d} = +0.015 \pm 0.032 \, {\mathrm (stat)} \pm 0.027 
\, {\mathrm (sys)} \, .
\end{equation}
This represents a factor of three improvement over the precision of the
Mainz  measurement. In the Standard Model, the dominant contribution
of Fig.~1A would result in
\begin{equation}
C_{2u} - C_{2d} = -(1-4\sin^2\theta_W) = -0.075 \, .
\end{equation}
Adding in the corrections associated with the amplitudes displayed
in Figs.~1B and 1C changes the value to $-0.058 \pm 0.02$.  Unless additional 
contributions to $G_A^e (T=1)$ are identified that can improve the
agreement, our measurement is in disagreement with the present
standard electroweak prediction at the 1.5 $\sigma$ level. 

An improved determination of $G^e_A$ will be
provided with a new measurement of parity-violating 
quasielastic electron-deuteron scattering at a lower beam energy.
The spatial dependence of the strange quark contributions and $G_A^e$
will be studied in detail in an upcoming program of parity violation 
measurements at the Thomas Jefferson National Accelerator Facility.

\newpage
\vskip 12pt
\centerline{\Large\bf References}
\vskip 12pt


\begin{enumerate}
\item G.~Miller, {\it et al.}, {\it Phys.~Rev.}~{\bf D5}, 528 (1972).

\item D.~Kaplan, A.~Manohar, {\it Nucl.~Phys.}~{\bf B310}, 527 (1988).

\item R.~D.~McKeown, {\it Phys. Lett.}~{\bf B219}, 140 (1989).

\item B.~A.~Mueller, {\it et al.}, {\it Phys.~Rev.~Lett.}~{\bf 78}, 3824 (1997).

\item D.~T.~Spayde {\it et al.}, {\it Phys.~Rev.~Lett.}~{\bf 84}, 1106 (2000).

\item Particle Data Group, D.~E.~Groom, {\it et al.},
{\it Eur.~Phys.~J.} {\bf C15}, 1 (2000).


\item I.~Zel'dovich, {\it JETP Lett.}~{\bf 33}, 1531  (1957).

\item Note that our definition of $F_A$ differs 
from that used in the atomic physics literature by a factor of $m^2 G_F$ 
(where $G_F$=1.16639(2)$\times$10$^{-5}$ GeV$^{-2}$ is the muon decay constant
and $m$ is the proton mass in units of (GeV/c$^2$)), 
with the result that we expect the value of $F_A$ to be of order unity.

\item C.~S.~Wood, {\it et al.,}, {\it Science} {\bf 275} 1759 (1997).
See also W.~Haxton, {\it Science} {\bf 275} 1753 (1997).

\item W.C.~Haxton, E.M.~Henley, M.J.~Musolf,
{\it Phys.~Rev.~Lett.}~{\bf 63}, 949 (1989).

\item  The contributions to the radiative
corrections in parity-violating electron scattering ($R^e$) compared to
neutrino scattering ($R^{\nu}$) are not quite the same, 
but the differences are calculated to be small compared to
our measurement. Calculations were presented in ref.~{\it 16} below,
but anapole-type contributions are not specifically identified.
For a breakdown into anapole contributions compared with other terms, 
see M.~Musolf, Ph.D. thesis, Princeton University, 1989.

\item  The authors here, and those in (4) and (5), constitute the SAMPLE 
collaboration. As stated in the text, the apparatus used here was identical 
to that used in (4) and (5) with the exception of the target fluid and some 
supplementary neutron shielding.


%
%

\item T.~Oka, 
{\it Phys.~Rev.}~{\bf D27}, 523 (1983).
 
\item E.~Hadjimichael, G.~I.~Poulis, T.~W.~Donnelly, 
{\it Phys.~Rev.}~{\bf C45}, 2666 (1992).

\item T.~R.~Hemmert, U.-G.~Meissner, S.~Steininger, {\it Phys.~Lett.}~{\bf B437},
184 (1998).

\item A.~Liesenfeld, {\it et al.}, {\it Phys.~Lett.}~{\bf B468}, 
19 (1999).

\item M. J. Musolf, B. R. Holstein, {\it Phys.~Lett.}~{\bf B242}, 461 (1990).

\item S.-L.~Zhu, {\it et al}., {\it Phys.~Rev.}~{\bf D62}, 033008 (2000).

\item C.~Y.~Prescott {\it et al.}, {\it Phys.~Lett.}~{\bf B84},
524 (1979).

\item W.~Heil {\it et al.}, {\it Nucl.~Phys.}~{\bf B327}, 1 (1989).



\item The skillful efforts of the staff of 
the MIT/Bates facility to provide
high quality beam and improve the experiment are 
gratefully acknowledged. Supported by
NSF grants PHY-9420470 (Caltech), PHY-9420787 (Univ.~of Illinois), 
PHY-9457906/PHY-9971819 (Univ.~of Maryland), PHY-9733772 (Virginia Polytechnic
Institute) and DOE cooperative agreement DE-FC02-94ER40818 (MIT-Bates) and
contract W-31-109-ENG-38 (Argonne National Laboratory), 
and the Jeffress Memorial Trust (William and Mary, grant no.~J-503).

\end{enumerate}

\newpage
\begin{figure}
\begin{center}
\epsfig{file=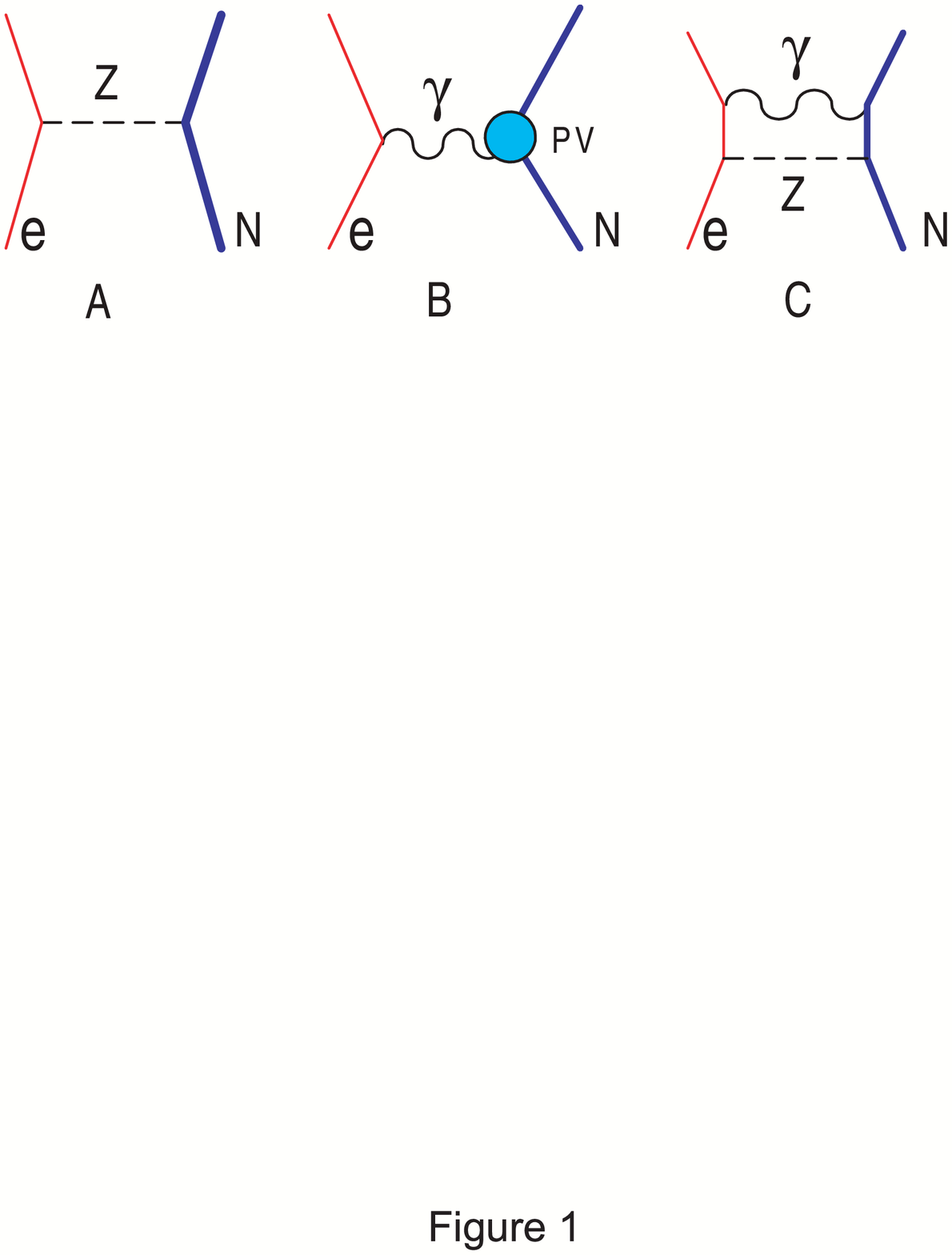, width=6in}
\vskip 0.6 in
\caption{Feynman diagrams representing three contributions to the axial
$e$-$N$ coupling: A) single $Z$-exchange, B) parity violating 
photon exchange, which contributes to the nucleon's anapole moment, 
and C) a $\gamma$-$Z$ box diagram typical of radiative corrections.}
\end{center}
\end{figure}

\begin{figure}
\begin{center}
\epsfig{file=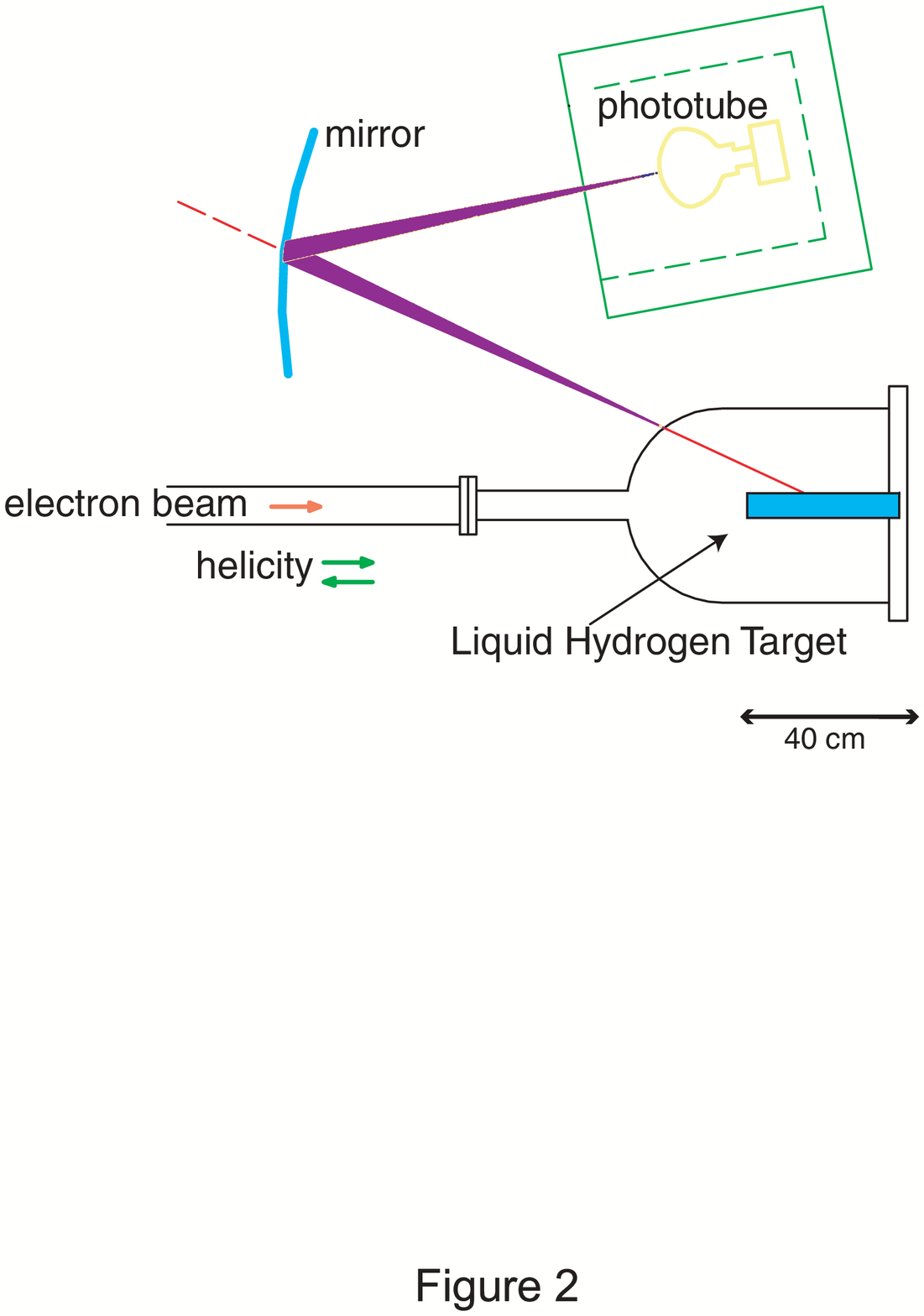,width=6in}
\vskip 0.2 in
\caption{A schematic view of one module of the SAMPLE 
experimental apparatus. Ten mirror-phototube pairs are placed
symmetrically about the beam axis.}
\end{center}
\end{figure}

\begin{figure}
\begin{center}
\epsfig{file=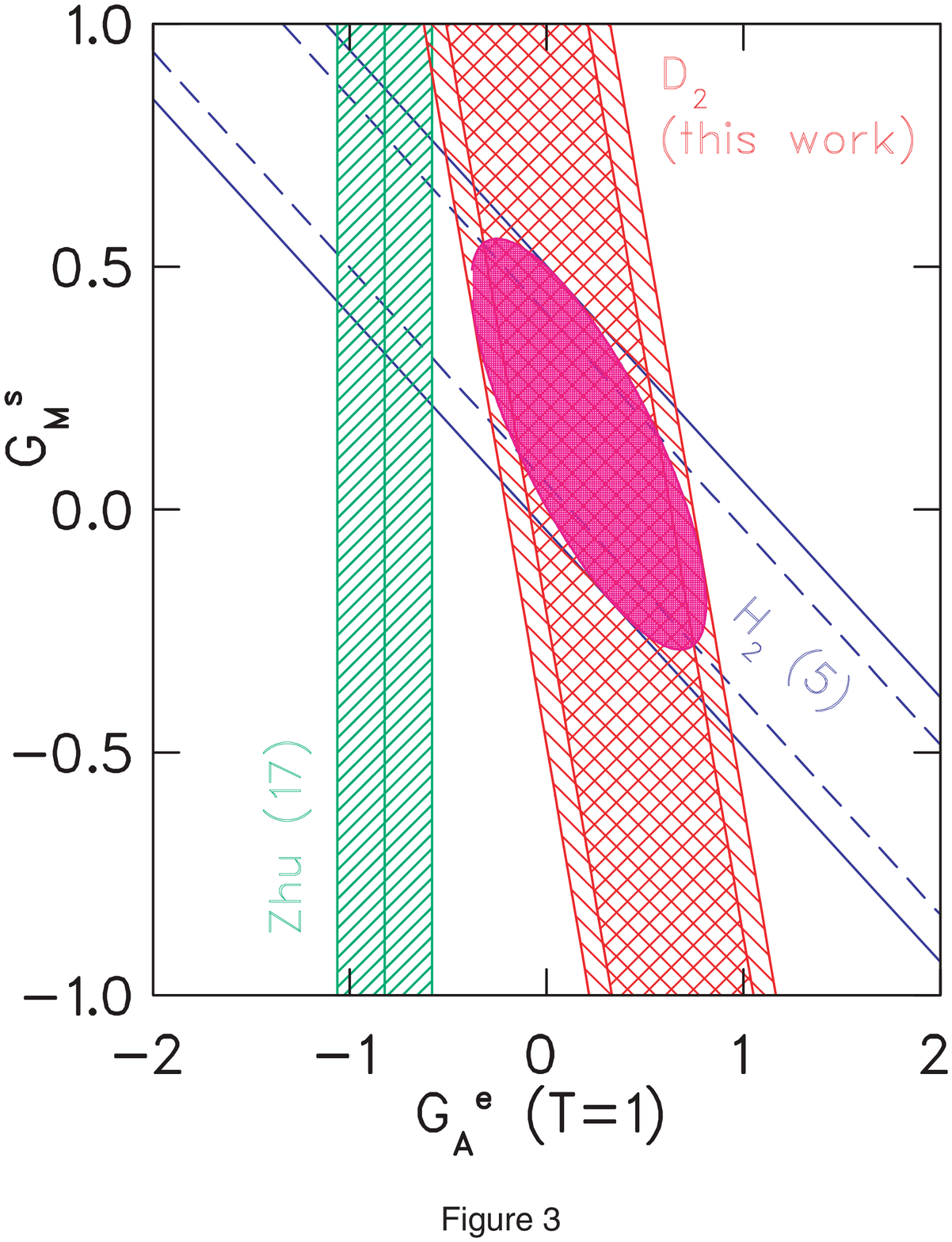,width=6in}
\vskip 0.2 in
\caption{A result of a combined analysis of the data from the previous
two SAMPLE measurements. The two error bands from the hydrogen
experiment~{\it (5)} 
and the  deuterium
experiment are indicated. The inner hatched region includes the
statistical error and the outer represents the systematic uncertainty
added in quadrature.  The ellipse represents the allowed region for
both form factors at the 1$\sigma$ level. Also plotted is the
estimate of the isovector axial $e$-$N$ form factor $G_A^e(T=1)$
obtained by using the anapole form factor and radiative corrections of
Zhu, {\it et al.}~{\it (13)}.
}
\end{center}
\end{figure}



\end{document}